\documentclass{sfb676}

\begin{document}
%------------------------------------
%\title{C7 --- Thermodynamics of Quantum Fields in Nonstationary Spacetimes}
\title{Thermodynamics of Quantum Fields in \\ Nonstationary Spacetimes}

%for single authors the superscripts are optional
\author{{\slshape Klaus Fredenhagen$^1$, Thomas-Paul Hack$^2$, Nicola Pinamonti$^3$}\\[1ex]
$^1$II. Institut f\"ur Theoretische Physik, Universit\"at Hamburg, Germany\\
$^2$Institut f\"ur Theoretische Physik, Universit\"at Leipzig, Germany\\
$^3$Department of Mathematics, University of Genova, Italy and INFN, Sez. Genova, Italy\\
}

% please enter the contribution ID for the DOI
\contribID{C7}

\desyproc{PUBDB-2018-00782}
\acronym{SFB 676 -- Particles, Strings and the Early Universe} 
\doi  % DOIs will bew registered by the DESY library

\maketitle

\begin{abstract}
Quantum field theory (QFT) on non-stationary spacetimes is well understood from the 
side of the algebra of observables. 
The state space, however, is largely unexplored, due to the 
non-existence of distinguished states (vacuum, scattering states, thermal states). 
Project C7 of the SFB 676 was focused on characterisations of states by 
asymptotic conditions, e.g. holography (in case the boundary has sufficiently 
many symmetries), on a precise version of an approximate particle interpretation 
(for instance in Robertson--Walker spacetimes) and on the 
determination in terms of expectation values of locally covariant fields. 
Additionally, the backreaction of quantum matter fields on the curvature as well as the  perturbative quantisation of the Einstein--Klein--Gordon system in the case of a cosmological background have been investigated. Finally, a detailed analysis and construction of equilibrium states for interacting field theories has been performed.
%
%
%
%Place your abstract here. It should not exceed 100 words.
%Please don't use footnotes in the abstract or title.
\end{abstract}

\section{Introduction}

Quantum field theory and General Relativity are two of the most successful physical theories ever developed. 
Their combination in a unified body is however still problematic. 
In spite of the many attempts, no universally accepted theory of quantum gravity is available in the literature.  In the recent years many interesting features of an eventual theory of quantum gravity have been discovered by analysing the unification of the two theories in approximated regimes like semiclassical gravity.
In quantum field theory on curved spacetime, gravity is still described by a classical curved spacetime while   
 matter is modeled by quantum fields which propagate on that classical spacetime.
In first approximation, their backreaction can be taken into account by means of the semiclassical Einstein equation 
\[
G_{ab}= 8\pi\langle T_{ab} \rangle_\omega
\]
which is formally similar to the ordinary Einstein equation where the stress tensor of the classical matter is substituted with the expectation value of the stress tensor of quantum matter in a suitable state $\omega$.  This equation is meaningful only if the fluctuations of the stress tensor are small. Interesting and well known effects like Hawking radiation for black holes or particle creation in cosmology can be obtained in this approximation. 

A physical scenario where these ideas can be tested is cosmology. 
Due to the homogeneity and isotropy hypothesis applicable in this case, the manifold describing the spacetime is $M=I\times \Sigma$ where $I$ is a real interval representing time and $\Sigma$ is the three dimensional manifold of space.
The geometry is described by the Friedmann--Lema\^itre--Robertson--Walker (FLRW) metric
\begin{equation}\label{eq:FLRWmetric}
ds^2 = - dt^2 + a(t)^2    \left(\frac{1}{1-\kappa r^2} dr^2  + r^2d\theta^2 + r^2\sin\theta^2 d\varphi^2  \right)
\end{equation}
where $\kappa\in\{-1,0,1\}$ distinguishes between open, flat or closed spatial sections and the scale factor $a(t)$ describes the way in which the universe evolves with respect to cosmological time. At present our spacetime appears to be almost spatially flat, for this reason we shall consider the case $\kappa=0$ only. Due to the time-dependence present in the metric, concepts like temperature, energy, particles and vacuum states cannot be used  -- one is forced to rethink all of them.

In order to fully understand the behavior of quantum matter in cosmological spacetimes, as part of the SFB Project we have analysed the construction of regular states and discussed some of their thermodynamical aspects~\cite{Dappiaggi:2007mx,Dappiaggi:2008dk,Dappiaggi:2010gt}. We have furthermore analysed the backreaction of quantum matter on curvature in the semiclassical approximation~\cite{Dappiaggi:2008mm,Dappiaggi:2009xj,Hack:2010iw}.
As a subsequent step, we have investigated the quantisation of the fluctuation of the matter-gravity system (Einstein--Klein--Gordon system) in the linear regime~\cite{Hack:2014epa} and beyond~\cite{Brunetti:2013maa}. 
We have discussed the form of gauge-invariant observables associated to this system when the background is chosen to be an FLRW spacetime~\cite{Brunetti:2016hgw}.
Finally, more recently, we have rigorously constructed equilibrium states for perturbatively constructed interacting field theories~\cite{Lindner:2013ila,Fredenhagen:2013cna}.

The project has been carried out using methods of Algebraic Quantum Field Theory.
In this approach, quantisation is formulated in two steps. The first consists in the determination of the set of observables and of the algebraic relations among them. In particular, the commutation relations of the theory are implemented at this level. The second steps consists in the analysis of the form of physically relevant states of the system. 
While the first step is well under control at least for the case of free theories, the choice of a physically relevant state is more difficult and cannot be done using local properties of the spacetime only. The problem of finding regular states also on cosmological spacetimes based on their global properties is not straightforward and was the first point of the analysis performed in this project. 

In the next chapter, we briefly outline the obtained results regarding the construction of regular states based on asymptotic spacetime properties and on the local notion of vacuum states.
In the third chapter we review the results obtained in this project concerning the backreaction of quantum matter on the curvature induced by the semiclassical Einstein equation. The forth chapter contains a brief recollection of results obtained regarding the  perturbative quantisation of the Einstein--Klein--Gordon system. Finally, we present the construction of equilibrium states for interacting field theory recently obtained in the framework of perturbative algebraic quantum field theory in the fifth chapter.

\section{Regular states on curved spacetimes}

In order to compute correlation functions on a curved spacetime and to analyse their thermal behavior it is necessary to have control on the state of the system. 
In particular, expectation values of local fields and their fluctuations need to be finite. 
This requirement constrains the possible quantum states and the class of Hadamard states satisfy this constraint~\cite{Kay:1988mu}. It is however not clear how to construct these states starting from initial conditions at some fixed time if the spacetime is not a-priori known.  

In this section we discuss various methods developed within this project to construct regular states for a free quantum scalar field $\phi$ on curved backgrounds $(M,g)$ which are globally hyperbolic spacetimes~\cite{Wald:1984rg}. Similar results hold for more general types of fields~\cite{Hack:2010iw}. 
The field we have in mind satisfies the equation
\[
P\phi := -\Box \phi + \xi R \phi + m^2\phi = 0
\]
where $\Box=\nabla_a\nabla^a$ is the d'Alembert operator, $\xi$ is the coupling to the scalar curvature $R$ and
$m$ is the mass of the field. The $*-$algebra of observables $\mathcal{A}$ associated to this field is generated by ``smeared fields'' $\phi(f)$ where $f\in\mathcal{D}(M)$, the set of compactly supported smooth functions. The product is such that it satisfies the canonical commutation relations 
\[
[\phi(f), \phi(h)] = i\Delta (f,h)
\]
where $\Delta=\Delta_R-\Delta_A$ is the causal propagator of the theory which is equal to the retarded-minus-advanced fundamental solution of $P$, both of which are characterised by $P\Delta_{R/A}(f)=f$, $\text{supp} \Delta_{R/A}(f) \subset J^{+/-}(\text{supp} f)$. We recall that the fundamental solutions exist and are unique on globally hyperbolic spacetimes, see e.g.~\cite{2008arXiv0806.1036B}. Since the field is uncharged, the involution $*$ coincides with the complex conjugation 
\[
\phi(f)^* = \phi(\overline{f}). 
\]
The algebra $\mathcal{A}$ is called on-shell if $\phi(Pf)=0$ for every $f\in\mathcal{D}(M)$. If this is not the case, the algebra is said to be off-shell. In the latter case, the on-shell projection can be realised by restricting the physically relevant states of the system to those which satisfy the equation of motion
\[
\omega(\phi(f_1)\dots \phi(Pg) \dots \phi(f_n))=0.
\]
In the case of a free scalar field, a state $\omega$ is characterised by its $n-$point functions
which are elements of $\mathcal{D}'(M^n)$, the set of distributions on smooth compactly supported functions, defined as
\[
\omega_n(x_1,\dots, x_n):=\omega(\phi(x_1) \dots \phi(x_n)).
\]
For Gaussian states, also called quasi-free states, the $n-$point functions with odd $n$ vanish while those with even $n$ are obtained from the two-point function as
\[
\omega_{2n}(x_1,\dots, x_{2n}) = \sum_{i\in P_{2n}}\omega_2(x_{i_1},x_{i_2})\dots \omega_2(x_{i_{2n-1}},x_{i_{2n}})
\]
where the sum is taken over all possible ordered sequences $P_{2n}$ of $\{1,\dots,2n\}$. A sequence $i$ is said to be ordered and thus contained in  $P_n$ if $i_{2k-1}<i_{2k}$ and $i_{2k'-3}<i_{2k'-1}$ for $k\in \{1,\dots,n\}$ and $k'\in \{2,\dots,n\}$.

Here we shall concentrate our attention on Gaussian states and thus we have to care only about the two-point function 
\[
\omega_2(x_1,x_2) = \omega(\phi(x_1)\phi(x_2)).
\]
As said before, the states we are considering need to be sufficiently regular to admit an extension to the algebra of local Wick polynomials which are pointlike products of fields and are realised extending the test functions $(f_1,\dots, f_n)\in \prod C^\infty_0(M)$ to distributions supported on the diagonal in $M^n$. Furthermore, the fluctuations of these fields need to be finite. In order to have these properties under control a sufficient condition is to require the state to be of Hadamard form.

A state of a free scalar field theory is of Hadamard form if on every normal neighborhood its two-point function can be expanded in the following way
\begin{equation}\label{eq:Hadamard}
\omega_2(x_1,x_2) = \lim_{\epsilon\to 0}\frac{U(x_1,x_2)}{\sigma_\epsilon(x_1,x_2)} + V(x_1,x_2)\log\sigma_\epsilon(x_1,x_2) + W(x_1,x_2)
\end{equation}
where $U,V$ are smooth functions on $M^2$ which are universal and depend only on the geometry and on the parameters present in the equation of motion, while $W$ is a smooth function which characterises the state. Furthermore, $\sigma_\epsilon(x_1,x_2) = \sigma(x_1,x_2) +i\epsilon(T(x_1)-T(x_2))$  
and $\sigma$ is one half of the square of the geodesic distance taken with sign while  $T$ is any time function.  

It is important to notice that Hadamard states are equivalently characterised by their wave front set. The wave front set of a distribution encodes its singular structure.
In particular, it is a subset of the cotangent space whose projection on the spacetime gives the singular support of the distribution while the points in the fiber determine the directions of non-rapid decrease of the Fourier transform of the distribution localised in any neighborhood of the corresponding base point. The precise definition can be found in the book of H\"ormander and it is used in the celebrated H\"ormander criterion of multiplication of distributions.  

The Hadamard state satisfies the ``microlocal spectrum condition'', namely, its 
wave front set is such that
\begin{equation}\label{eq:muc}
WF(\omega_2)  = \{ (x_1,x_2;k_1,k_2)\in T^*{M^2}\setminus \{0\} | (x_1,k_1)\sim(x_2,-k_2), k_1\triangleright 0  \}
\end{equation}
where $(x_1,k_1)\sim(x_2,-k_2)$ if $x_1$ and $x_2$ are joined by a null geodesic and $g^{-1}k_1$ and $-g^{-1}k_2$ are the corresponding tangent vector with respect to an affine parametrisation. Furthermore,  $k_1\triangleright 0$ if $k_1$ is future directed. 
Radzikowski proved in~\cite{Radzikowski:1996pa} that a state is of Hadamard form if and only if it satisfies the microlocal spectrum condition. A generalisation of the microlocal spectrum condition to higher order Wick polynomials can be found in~\cite{Brunetti:1995rf}.
The microlocal spectrum condition can be understood as a covariant remnant of the spectrum condition usually employed in the case of vacuum states for quantum theories on flat spacetimes. It is a remarkable fact that this covariant remnant fixes the singular structure of the two-point function.  

The characterisation of Hadamard states by means of the wave front set of the two-point function is quite useful. It is actually much easier to check if a state satisfies the microlocal spectrum condition, then to precisely control the singular structure of its two-point function. Standard results of microlocal analysis like H\"ormander's propagation of singularity theorems can in fact be successfully employed.

\subsection{States from asymptotic properties}\label{se:asymptotic}

As mentioned above it is not straightforward to get Hadamard states out of initial values if the spacetime is not known a priori.
The problem is simpler when the initial values are given on null surface (characteristic initial value problem). In fact, in this situation the positive frequency condition present in the microlocal spectrum condition can be tested directly for the initial values.
This is the basic idea at the heart of the construction of Hadamard states based on asymptotic initial values when the asymptotic past infinity (or more generally the asymptotic past boundary) is a causal cone. In the following we discuss these basic ideas in a more rigorous form.
Consider a spacetime $M$ whose past boundary is a null cone $\mathcal{C}$ in another  (conformally related) larger spacetime $\tilde{M}$. Due to the well-posedness of the Goursat problem, it is known that giving initial values on $\mathcal{C}$ fixes uniquely a solution of any hyperbolic equation in $M$.

As an application of the time slice axiom~\cite{2009CMaPh.287..513C}, we may thus construct a map from the on-shell algebra $\mathcal{A}(M) \to \mathcal{A}(C)$ which is an injective $*-$homomorphism. This is essentially done at the level of generators of the algebra $\mathcal{A}(M)$ and 
\[
\iota \phi(f) = \Psi(f_{\mathcal{C}}), \qquad   f_{\mathcal{C}} = \left. \Delta (f) \right|_{\mathcal{C}}
\]
where the causal propagator $\Delta$ is seen in $\tilde{M}$ and  $\Psi(f)$ are the generators of $\mathcal{A}(C)$.
On $C$ there is a notion of causality, it is thus possible to construct states which look like a vacuum, meaning that their spectrum contains only positive frequencies with respect to causal directions on $\mathcal{C}$.
If we indicate such a state by $\Omega^+$, it turns out that its pullback 
\[
\omega(A) = \iota^*\Omega^+(A) = \Omega^+(\iota(A))
\]
defines a state in $M$ which is an Hadamard state.

This idea has been used in various physical situation in order to determine the singular structure of known states:
\begin{itemize}
\item in the case of asymptotically flat spacetime to check the Hadamard property of asymptotically vacuum states for massless particles~\cite{Dappiaggi:2005ci};
\item in the case of asymptotically de Sitter spacetimes~\cite{Dappiaggi:2007mx,Dappiaggi:2008dk} to discuss the Hadamard properties of Bunch--Davies-like states;
\item in the case of spacetimes possessing a null big bang scenario to control asymptotically vacuum states~\cite{Pinamonti:2010is};
\item in the case of Schwarzschild spacetime to prove that the Unruh state is of Hadamard form~\cite{Dappiaggi:2009fx}.
\end{itemize}

For a cosmological spacetime $(M,g)$ with $g$ as in \eqref{eq:FLRWmetric}, pure Gaussian states which are invariant under the spatial isometries of the metric have a two-point function of the form:
\begin{equation}\label{eq:two--point-function}
\omega_2(x_1,x_2) :=  \lim_{\epsilon\to 0^+}\frac{1}{8\pi^3a(t_1)a(t_2)} \int_{\mathbb{R}^3}  \chi_k(t_1)\overline{\chi_k(t_2)}  e^{i{\mathbf{k}}\cdot (\mathbf{x}_1-\mathbf{x}_2)}   e^{- k \epsilon}     d\mathbf{k}
\end{equation}
whereby $k = |\mathbf{k}|$ and the mode functions $\chi_k(\tau)$ satisfy the equation
\[
\chi_k''+(m^2 a^2 +k^2)\chi_k + \left(6\xi-1\right) \frac{a''}{a}\chi_k =0,
\]
where $'$ denotes the derivative with respect to conformal time $\tau$ obtained by integrating the relation $dt = a d\tau$, and the modes are normalised such that the following Wronskian condition
\[
\overline{\chi_k}'\chi_k - 
\overline{\chi_k}\chi_k' =i \;
\]
holds. The latter condition ensures that the antisymmetric part of $\omega_2$ is proportional to the causal propagator of the theory~\cite{Lueders:1990np}.
As discussed above, it is not easy to find the modes $\chi_k$ for which $\omega_2$ as in \eqref{eq:two--point-function} correspond to an Hadamard state. 
The states which are asymptotic vacuum states constructed with the pullback of $\iota$ discussed above, are usually quasi-free states of the from \eqref{eq:two--point-function} and the modes have the asymptotic form
\[
\chi_k(\tau) \underset{{\tau\to-\infty}}{\longrightarrow} \frac{e^{ik\tau}}{\sqrt{2k}}.
\] 
The construction of asymptotic vacuum states can be generalised to construct approximate KMS states. In particular, if the boundary state is a KMS state with respect to the characteristic translation, the obtained state in the bulk can be interpreted as an approximate KMS state and its two-point function is 
\begin{equation}\label{eq_genth}
\omega_2(x_1,x_2) :=  \lim_{\epsilon\to 0^+}\frac{1}{8\pi^3a(t_1)a(t_2)}
\int_{\mathbb{R}^3}  \left(\frac{\overline{\chi_k(t_1)}\chi_k(t_2)}{1-e^{-\beta k_0}}+\frac{\chi_k(t_1)\overline{\chi_k(t_2)}}{e^{\beta k_0}-1}\right)e^{i{\mathbf{k}}\cdot ({\mathbf{x}}_1-{\mathbf{x}}_2)}   e^{- k \epsilon}     d\mathbf{k}
\end{equation}
with $k_0:=|\mathbf{k}|$. Other approximate KMS states can now be constructed considering 
$k_0:=\sqrt{k^2+m^2a^2_F}$ for a fixed constant $a_F$. Further comments on this can be found in~\cite{Dappiaggi:2010gt}.

\subsection{States of low energy}

Olbermann proposed in~\cite{Olbermann:2007gn} a way to select the mode functions $\chi_k$ in \eqref{eq:two--point-function} in order to obtain a state which satisfies the microlocal spectrum condition. The idea
is to find the modes which minimise the energy density smeared in a chosen smooth way along a fixed time interval. 

For this reason these states are called states of low energy.  Although they depend on the chosen smooth smearing function they can be interpreted as approximate vacuum states. Excitations of these states are natural candidates for approximate particle states, and first results have already been obtained in~\cite{Degner:2009rq}. An interesting observation is that the number of particles is 
oscillating during the cosmological evolution. Whether this is an artefact of 
the approximation or a genuine physical effect has to be clarified. 

Within this project, a further extension of these ideas to states which are locally in equilibrium has been discussed in the PhD thesis of K\"usk\"u~\cite{Kusku:2008zz}.

\subsection{Sorkin--Johnston states and generalisations thereof}

On a static spacetime, the two-point function of the vacuum state of the theory can be obtained from the spectral decomposition of the causal propagator and its restriction to the positive part of the spectrum. 
A local version of this construction was proposed by Sorkin--Johnston, see~\cite{Afshordi:2012jf} and reference therein. The basic difference between their idea and the above is to construct a local vacuum starting from the spectral decomposition of the causal propagator localised in some double cone region. Unfortunately, in general, the states obtained in this way do not satisfy the microlocal spectrum condition \cite{Fewster:2012ew}. 

Within the present project, Brum and Fredenhagen in~\cite{Brum:2013bia} showed that a smoothly smeared version of the Sorkin--Johnston construction results in a state which is of Hadamard form. 
To construct these states, one mollifies the sharp localisation of the causal propagator to a double cone $N\subset M$ by choosing a real-valued $f\in \mathcal{D}(M)$ such that $f$ is equal to $1$ on $N$. This function is then used to construct the operator
\[
A:=if \Delta f
\]
which can be extended to act on the Hilbert space $L^2(M, d\mu_g)$. The operator obtained in this way is, at least on static and on FLRW spacetimes, bounded and self-adjoint, hence, by means of standard functional calculus, it is possible to extract its positive part
\[
A^+ := \Pi^+A
\]
where $\Pi^+$ is the spectral projection on the interval $[0,|A|]$. 
The generalised Sorkin--Johnston state is now obtained by considering the quasifree state characterised by the two-point function associated to $A^+$:
\[
\omega_2(h_1,h_2):=\left(h_1,A^+ h_2\right), \qquad h_i\in\mathcal{D}(N).
\]
The state obtained in this way turns out to be a Hadamard state.

\section{Semiclassical gravity}

At first approximation, the influence of quantum matter on the curvature can be taken into account by means of the semiclassical Einstein equation
\[
G_{ab} = \langle T_{ab} \rangle_\omega
\]
whereby gravity is described by a classical curved spacetime. This equation is formally similar to the classical Einstein equation where the stress tensor of the classical matter is substituted with the expectation value of a the stress-tensor of some quantum field propagating on a curved background.

In the cosmological case, if the conservation equation of the stress tensor
\[
\nabla^a \langle T_{ab} \rangle_\omega = 0
\]
holds, the semiclassical Einstein equation is equivalent to 
\[
-{R_a}^a = 8\pi \langle {T_a}^a \rangle_\omega
\]
up to an initial condition which fixes the amount of radiation present in the solution. 
In the expectation value $T=\langle{T_a}^a \rangle_\omega$ of the trace of the stress tensor we can recognize three contributions
\[
-{R_a}^a    = 8\pi \left( T_{state} + T_{anomalous}+  T_{ren.freed.}    \right)
\]
corresponding to a state dependent part, an anomalous contribution due to the requirement of covariant conservation, see e.g.~\cite{PhysRevD.17.1477,Moretti:2001qh,Hollands:2004yh}, and a contribution which can be interpreted as renormalisation freedom~\cite{Hollands:2001nf}.  
In the case of $\xi=1/6$, the anomalous contribution is
\[
T_{anomalous} =   \frac{1}{8\pi^2} \left( \frac{1}{360} \left( C_{abcd}C^{abcd}+R_{ab}R^{ab}-\frac{R^2}{3}+\Box R \right) + \frac{m^4}{4} \right)
\]
with the Weyl tensor $C_{abcd}$.
The renormalisation freedom is 
\[
T_{ren.freed.} =   \alpha  m^4 +\beta m^2 R +\gamma \Box R,
\]
whereas the state dependent contribution is 
\[
T_{state} =   \frac{1}{8\pi^2} m^2  [W]
\]
with $[W]$ indicating the coinciding point limit of the smooth part of the state's two-point function \eqref{eq:Hadamard}.
With the choice $\gamma=0$ and a suitable assumption on the state dependent contribution (considering it to be vanishing like in zeroth-order adiabatic states) and in the case of massive, conformally coupled fields, the equation for the trace is an ordinary differential equation 
\begin{equation}\label{eq:diff}
\dot{H}(H^2-H_0^2)=-H^4+2H^2H_0^2+M;
\end{equation}
where $H=\dot{a}/a$ is the Hubble parameter and $\dot{\bullet}$ denotes a derivative with respect to cosmological time. The constants
\[
H_0^2 = 180\pi\left(\frac{1}{G} - 8\pi m^2\beta\right), 
\qquad 
M = \frac{15}{2}m^4-240\pi m^4 \alpha 
\]
depend on the renormalisation freedom. $\beta$ can be interpreted as a renormalisation of the Newton constant and for this reason it can be fixed to be $0$, $\alpha$ represents a renormalisation of the cosmological constant and can be set to model the present value of the measured constant. 

The solution space of the differential equation \eqref{eq:diff} can be easily obtained. We notice in particular that there are two stable solutions, the first of which could describe the phase of rapid expansion present in the beginning of the universe, whereas the second one can model the present universe and its recent past.
However, in this simplified model there is no mechanism which permits to exit the phase of rapid expansion. The oversimplifications present in the model are the following: the considered matter field is a free field, the role of higher derivatives which are present in the equation for a different choice of $\gamma$ could have an important role and the state dependent contribution is not taken into account properly. 

The results presented in this section are summarised in the paper~\cite{Dappiaggi:2008mm}. Within this project it has been shown that similar results hold also when the matter is described by other fields like Dirac fields~\cite{Hack:2010iw}. In particular, the trace anomaly, the form of stress tensor and the extended algebra of fields for Dirac fields has been analyzed in~\cite{Dappiaggi:2009xj}.

\subsection{Rigorous analysis of the semiclassical equation}

The state dependent contribution present in the trace makes the dynamical problem much more complicated, essentially because it is difficult to fix the state with a  universal prescription that does not depend on the particular spacetime. 

This problem has been rigorously studied in two cases. The first case considered were  cosmological spacetimes which possess an initial null big bang phase. In fact, when the past boundary of $M$ is a null surface, it is possible to use similar techniques like the ones discussed above in section \ref{se:asymptotic}. In this way it has been proved in 
\cite{Pinamonti:2010is} that the semiclassical Einstein equation leads to a well-posed initial value problem. This means that solutions exists and are unique, at least for small amounts of time after the null big bang. 

Relaxing the request of having full Hadamard states, but considering adiabatic states of order $0$ as given in~\cite{Lueders:1990np} at a fixed time $t_0$, it was possible to extend the proof of existence also to this situation. The corresponding results have been summarised in~\cite{Pinamonti:2013wya}, where it is proved that solutions of the semiclassical Einstein equation exist and are unique up to future infinity or up to the occurrence of a future blowup of the scale factor (big rip).

\section{Quantum gravity and cosmological fluctuations}

To go beyond the semiclassical approximation presented above, and to quantise the degrees of freedom of both matter and gravity fields, we have analyzed the quantisation of the fluctuations of the coupled Einstein--Klein--Gordon system, namely, a massive minimally coupled scalar field $\phi$ propagating over a Lorentzian spacetime $(M,g)$ for which
\begin{equation}\label{eq:EKG}
R_{ab}-\frac{1}{2}R g_{ab} = 8\pi  T_{ab}    ,\qquad     - \Box \phi + m^2 \phi + V'(\phi)= 0,
\end{equation}
where $T_{ab}$ are the components of the stress tensor of $\phi$.

To discuss the linearised perturbation of the theory we consider a field configuration $\Gamma=(g,\phi)$ and expand it around a background solution $\Gamma_0$

\[
\Gamma = (g,\phi) = \Gamma_0 + \lambda   (\gamma,\varphi) + O(\lambda^2),
\]
equations for $(\gamma,\varphi)$ can now be obtained from \eqref{eq:EKG}.

In the case of perturbations in inflationary cosmological models, the background theory corresponds to an FLRW  spacetime plus an  inflaton $\phi$ which is constant in space. 

The linearised perturbations $(\gamma,\varphi)$ can be quantised using standard methods of algebraic quantum field theory on curved background in a manifestly covariant and gauge-invariant manner. Details of this construction are presented in the paper~\cite{Hack:2014epa}. In particular, a non-local behavior shown by the quantum Bardeen potentials is obtained and it is furthermore shown that a similar effect is present in any local quantum field theory. Finally the known linearised equation for the Mukhanov--Sasaki variable is recovered in that formalism.

\subsection{Beyond the linearised approximation}

The perturbative construction of quantum gravity in a generally covariant way has been presented in~\cite{Brunetti:2013maa}. This construction has been performed in the framework of locally covariant quantum field theory and the gauge problem has been treated by means of the renormalised Batalin--Vilkovisky formalism~\cite{Rejzner:2011jsa}. 
The problem of non-renormalisability of the theory persists and, hence, the theory is interpreted as an effective theory at large scales. In spite of this fact, the construction leads to a background-independent theory. Moreover, gauge-invariant observables are considered to be relational observables. This means that they are functionals of four scalar coordinate fields constructed which can be chosen as suitable curvature scalars, see also~\cite{Khavkine:2015fwa} for related work on these functionals. This idea can be used to construct gauge-invariant observables in cosmology in a simple way.

In the case of the Einstein--Klein--Gordon system the procedure works as follows. One selects 4 scalar fields $X_{\Gamma}^a, a=1,\ldots 4$, which are local functionals of the field configuration $\Gamma$ which contains the spacetime metric $g$, the inflaton field $\phi$ and possibly other fields. The fields $X_{\Gamma}^a$ are supposed to transform under diffeomorphisms $\chi$ as           
\begin{equation}\label{equivariance}
X_{\chi^*\Gamma}^a=X_{\Gamma}^a\circ\chi\,.
\end{equation}
We choose a background $\Gamma_0$ such that the map
\begin{equation}
X_{\Gamma_0}:x\mapsto (X_{\Gamma_0}^1,\ldots,X_{\Gamma_0}^4)
\end{equation}
is injective. We then set for $\Gamma$ sufficiently near to $\Gamma_0$
\begin{equation}
\alpha_{\Gamma}=X_{\Gamma}^{-1}\circ X_{\Gamma_0}.
\end{equation}
We observe that $\alpha_{\Gamma}$ transforms under diffeomorphisms as
\begin{equation}
\alpha_{\chi^*\Gamma}=\chi^{-1}\circ\alpha_{\Gamma}\,.
\end{equation}
Let now $A_{\Gamma}$ be any other scalar field which is a local functional of $\Gamma$ and transforms under diffeomorphisms as in \eqref{equivariance}. Then the field
\begin{equation}
\mathcal {A}_{\Gamma}:=A_{\Gamma}\circ\alpha_{\Gamma}
\end{equation}
is invariant under diffeomorphisms and may be considered as a local observable. Note that invariance is obtained by shifting the argument of the field in a way which depends on the configuration.

In~\cite{Brunetti:2016hgw} this idea is used to show that the standard theory of cosmological perturbations arises as the linear order of a fully quantised perturbative theory of quantum gravity. In particular, in the way explained above, it was possible to obtain gauge-invariant local observables which can be used to generate perturbative invariant expressions at arbitrary order. 
The standard gauge-invariant observables of cosmological perturbation theory were obtained at first order and explicit expressions at second order were explicitly given. Due to the large degree of symmetry present in the case of cosmological backgrounds,  the spatial coordinate fields have been selected to be harmonic coordinates. While this choice works perfectly at the classical level, the gauge invariant observables constructed with them display severe non-localities, which could create problems in the 
quantisation of the system. Further investigations are needed in order to clarify these points. The ideas of these project have been recently developed further in~\cite{Frob:2017lnt}.

\section{Perturbative algebraic quantum field theory and equilibrium states}

Whenever an interacting theory is treated perturbatively, 
its elements are usually given as a formal series in the coupling constant.
Three major problems arises in these kind of treatments:
\begin{itemize}
\item there are ultraviolet problems at each order in perturbation theories;
\item there are infrared problems when the interaction Lagrangian is supported in a non compact domain;
\item the perturbative series does not converge.
\end{itemize}
The first problem is nowadays well understood. The second problem can be treated analysing the adiabatic limits of the theory. Regarding the third problem, very little is known. In some integrable cases, usually in lower dimensional theories, it is nevertheless possible to compare perturbatively obtained results with exact ones. As an example 
the Sine--Gordon model can be treated both exactly and perturbatively. In~\cite{Bahns:2016zqj} it is shown that the perturbative methods agrees with the exact treatment. Further developments can be found in~\cite{Bahns:2017mrt}.

During the development of the present project, methods of algebraic quantum field theory have been developed to treat non-linear theories perturbatively. In particular, the framework of perturbative algebraic quantum field theory (pAQFT) was developed in~\cite{Brunetti:2009qc}.
Using this framework, it was possible to clarify the connections between Wilson's concept of the renormalisation group 
and the original concept of St\"uckelberg and Petermann. Furthermore, it was recently possible to construct rigorously equilibrium states for interacting theories in the adiabatic limit in~\cite{Lindner:2013ila} and in~\cite{Fredenhagen:2013cna}. Central in this construction was the use of the time slice axiom proved in~\cite{2009CMaPh.287..513C}.

\subsection{Perturbative algebraic quantum field theory}

With the aim of discussing the construction of equilibrium (thermal) states for interacting theories, 
we briefly recall the main steps of pAQFT in the simple case of an interacting scalar field theory.
The starting point is a concrete realisation of the observables as functionals of field configurations
\[
\phi\in C^\infty(M;\mathbb{R}).
\]
In order to be able to work with them we require that the observables are smooth local functionals, namely, their functional derivatives exist to all orders and at each order they are described by compactly supported distributions. 
Furthermore, the singular structure of these distributions must be under control.  
Hence, the observables of the theory are represented by the set of microcausal functionals 
\[
	\mathcal{F}_{\mu c} = \left\{ F:\mathcal{C}\to\mathbb{C}  \left| \forall n \; F^{(n)}\in\mathcal{E}'(M^n) , \text{WF}({F^{(n)}})\cap (\overline{V}_+^n\cup\overline{V}_+^n)  =\emptyset  \right.\right\}.
\]
Among all possible elements of $\mathcal{F}_{\mu c}$ the regular ones  are those whose functional derivatives are smooth functions. They correspond to the elements of the algebra generated by linear fields smeared by smooth test functions. 
The set of regular functionals is denoted by $\mathcal{F}_{reg}$ and it does not contain Wick polynomials.
The local functionals are those whose functional derivatives are supported on the diagonals, they are denoted by $\mathcal{F}_{loc}$ and the Wick polynomials are contained in this set.
The free algebra of observables can be constructed equipping microcausal functionals with the following product
\[
F \star_\omega G = \mathcal{M} \exp{\left(\frac{1}{2} \Gamma_\omega \right)} F \otimes G
\]
where $\mathcal{M}$ denotes the pointwise multiplication $\mathcal{M}(F\otimes G) = FG$ and
\[
\Gamma_\omega = \int \omega_2(x,y) \frac{\delta}{\delta \phi(x)}\otimes\frac{\delta}{\delta \phi(y)}
\]
where $\omega_2$ is any two-point function of Hadamard form of the chosen free (linear) quantum theory. A two-point function of Hadamard form solves in particular the free equation of motion up to smooth functions, its antisymmetric part is proportional to the causal propagator of the free theory and its wave front set satisfies the microlocal spectrum condition \eqref{eq:muc}.

The product defined above is well-posed among functionals which possess only a finite number of non-vanishing functional derivatives. It can however be used also among more general functionals if a suitable weaker notion of convergence is considered. This is for example the case if observables are formal power series whose coefficients are elements of $\mathcal{F}_{\mu c}$.
When equipped with the involution defined by means of the complex conjugation as $F^*(\phi) = \overline {F(\phi)}$ the $(\mathcal{F}_{\mu c},\star_\omega,*)$ represent the $*-$algebra of the Wick polynomial normal ordered with respect of $\omega$. These are the observables of a free theory.

The canonical commutation relations are encoded in the form of the product $\star_\omega$, in fact, if one considers linear fields 
\[
F_f(\phi)=\int f \phi \,d\mu_g, \qquad f\in \mathcal{D}(M)
\]
it holds that
\[
[F_f,F_h]_{\star_\omega} = F_f\star_\omega F_h - F_h\star_\omega F_h = \omega(f,h) - \omega(h,f) = i\Delta(f,h)
\]
where $\Delta$ is the causal propagator of the free theory, namely, the advanced-minus-retarded fundamental solution. These commutation relations do not depend on the particular form of the chosen Hadamard two-point function $\omega_2$. 

Even more, giving two Hadamard functions $H,H'$, the map
\[
\alpha_{H-H'}: (\mathcal{F}_{\mu c},\star_{H'},*) \to(\mathcal{F}_{\mu c},\star_{H},*)
\]
defined by 
\[
\alpha_{w} F = \exp \frac{1}{2}\Gamma'_{w} F  , \qquad   \Gamma'_w = \int w(x,y) \frac{\delta^2}{\delta \phi(x)\phi(y)} d\mu_g(x)d\mu_g(y)
\]
is actually a $*-$isomorphism between the two algebras. 

If one restricts attention to the regular functionals, one obtains the well-known Borchers--Uhlmann algebra of fields.  We stress that the algebra obtained here is off-shell, this is an essential step in order to be able to treat interactions perturbatively.

The observables associated to interacting field theories can be constructed by means of perturbation theory. To this end one has to consider the time-ordered product.
This product satisfies certain requirements, a corresponding list of axioms can be found in~\cite{Hollands:2001nf}. 
The existence of time-ordered products of local Wick polynomials on curved spacetimes has been established in~\cite{Brunetti:1999jn} and a covariant generalisation of this construction can be found in~\cite{Hollands:2001fb}. 
These constructive methods are generalisations of the inductive construction of Epstein--Glaser~\cite{Epstein1973}, according to which, at each inductive step $n$,  
the causal factorisation property permits to construct the time-ordered product of $n+1$ local fields up to the total diagonal in terms of the time-ordered products of $n$ local fields.
The extension to the total diagonal of the obtained distributions can be done by means of Steinmann-scaling-degree~\cite{steinmann2014perturbation} techniques.
The obtained time-ordered product is a map from multilocal functionals to microcausal functionals
\[
T:\mathcal{F}_{\text{loc}}^{\otimes n} \to \mathcal{F}_{\mu c}
\]
and can be used to construct time-ordered exponentials (local $S$-matrices) of a local functional $V\in\mathcal{F}_{\text{loc}}$
\[
S(V)= \text{exp}_T (V). 
\]
This time-ordered exponential is used to represent the interacting local fields in the free algebra by means of the Bogoliubov-map
\[
R_V(F) 
%:=  \left.\frac{d}{d\lambda}S(V)^{-1}\star S(V+\lambda F) \right|_{\lambda=0}
:= \left.\frac{d}{d\lambda}S(V)^{-1}\star S(V+\lambda F) \right|_{\lambda=0}.
\] 
The causal factorisation property at the level of $S$-matrices is such that for every $F,H,G\in\mathcal{F}_{\text{loc}}$ with $F$ later than $G$ 
\[
S(F+G+H) = S(F+G) \star S(G)^{-1} \star S(G+H),
\]
whereby $F$ is said to be later then $G$  if $\text{supp} F \cap J^-(\text{supp} G) =\emptyset$. 

%
%Hence, interacting fields can be constructed by means of the Bogoliubov formula
The interacting algebra is considered to be the algebra generated by the images of local fields under the Bogoliubov map
\[
\mathcal{F}_I\subset \mathcal{F}_{\mu c}.
\]
States for the interacting algebra are characterised by their expectation values on $\mathcal{F}_I$. Hence, every state of the free theory can be promoted to a state for the interacting theory by simply composing it with the Bogoliubov-map (or, rather,  products of Bogoliubov-maps applied to the generators of the interacting algebra). However, the physical meaning of a given state changes. In particular, in the case of a Minkowski spacetime, the free vacuum and equilibrium states of the theory are not mapped to vacuum and equilibrium states of the interacting theory.
However, in the adiabatic limit, the vacuum of the interacting theory can be shown to be equivalent to the 
free vacuum in this sense, while this is not the case for equilibrium states at finite temperature.
%
%\begin{itemize}
%\item 
%
%\item We discuss now how to construct equilibrium states in a static globally hyperbolic spacetime 
%
%\end{itemize}

As part of this project, it has been proven that the {\bf time slice axiom} holds both for  free and for interacting theories~\cite{2009CMaPh.287..513C}.
This result says that if a state is fixed in an $\epsilon-$neighborhood of a Cauchy surface in a globally hyperbolic spacetime, then it is fixed everywhere.

This observation implies that, if we want to define a state for an interacting field theory, it is sufficient to restrict the attention to field observables supported in an $\epsilon-$neighborhood of a Cauchy surface $\Sigma$. Denoting by $T$ the time function associated to the Cauchy surface $\Sigma$, such that the various Cauchy surfaces are the level sets of $T$, we have
\[
\Sigma_\epsilon := \left\{ p \in M \left |\; |T(p)-T(p_0)|\leq \epsilon , p_0\in\Sigma\right.\right\}.
\]
Thanks to the time slice axiom, to fix a state, it is sufficient to prescribe the expectation values of elements of
\[
\mathcal{F}_{\mu c}(\Sigma_\epsilon) = \{ F\in\mathcal{F}_{\mu c}| \text{supp} F\subset \Sigma_\epsilon \}
\]
for free field theories, or of the subalgebra  $\mathcal{F}_{I}(\Sigma_\epsilon)\subset \mathcal{F}_{\mu c}$ generated by $\{R_V(F)|F\in \mathcal{F}_{\text{loc}}(\Sigma_\epsilon)\}$ 
in the case of interacting theories.

 The time slice axiom permits us to restrict the domain of interacting observables. However, although the domain of an interacting field $R_V(F)$ coincides with the domain of $F$, $\text{supp}R_V(F)$ is larger than $\text{supp} F$. In fact, causality implies that
\[
\text{supp}R_V(F) \subset J^{-}(\text{supp} F) \cap J^{+}(\text{supp} V). 
\]
Hence, in order to localize the elements of $\mathcal{F}_{I}(\Sigma_\epsilon)$, we need to control the support of $V$.
%
%if $V$ does not have compact support, $\mathcal{F}$
%
%is a the interac is still non local. 
%
This last step can be done using the causal factorisation property. For every $F$ supported in $\Sigma_\epsilon$, we have that 
\[
R_V(F) = R_{V'}(F)
\]
if $V-V'$ is supported in the future of $\Sigma_\epsilon$ and that 
\[
R_V(F) = U^{-1} \star R_{V'}(F) \star U
\]
if $V-V'$ is supported in the past of $\Sigma_\epsilon$, where
\[
U = S^{-1}(V')\star S(V)\,.
\]
If $V$ and $V'$ are local functionals $S^{-1}=S^*$ and thus  both $S$ and $U$ are unitary and the algebras generated by 
$R_V(\mathcal{F}_{\text{loc}})$ and $R_{V'}(\mathcal{F}_{\text{loc}})$ are $*-$isomorphic with the isomorphism being implemented by the unitary $U$.

With this in mind, restricting the attention to a Minkowski spacetime and considering the interaction Lagrangian
\[
V_\lambda=\int \lambda \mathcal{L}_I d\mu_g
\]
with $\lambda$ a smooth and compactly supported function, it is possible to analyze the adiabatic limit as the inductive limit
\[
R_{V}(F) =  \lim_{\lambda\to1} R_{V_\lambda}(F)\,.
\]
This limit is called algebraic adiabatic limit, further details can be found in~\cite{Brunetti:1999jn}.

In Minkowski spacetime we can use the spacetime symmetries to adapt the form of the cutoff $\lambda$ to the geometry of the background. Using a standard Minkowskian coordinate system $(t,\mathbf{x})$ where $t$ parametrises the time and $\mathbf{x}$ the space, we choose    
\[
\lambda(t,\mathbf{x}) = \chi(t) h(\mathbf{x})
\]
where $\chi$ is a smooth time cutoff which can be chosen to be supported in $J^+(\Sigma_{2\epsilon})$ and equal to $1$ in $J^+(\Sigma_{\epsilon})$.
The spatial cutoff $h$ is a smooth function which is constant in time and compactly supported in space. Using the time slice axiom and the causality of the $S$ matrix, up to suitable unitaries of the above mentioned type, 
$\chi$ can be fixed once and for all and the adiabatic limit amounts to $h\to1$. 
For every $F\in \mathcal{F}_I(\mathcal{O})$, the algebraic adiabatic limit $h\to 1$ converges, in fact if $h_n$ is a family of compactly supported smooth functions which are equal to $1$ on the sphere $S_{n} = \{\mathbf{x}\in\mathbb{R}^3 | |x| < n \}$, due to causality, the sequence 
$R_{V_{\chi h_n}}(F)$ is constant for $n\geq \overline{n}$ depending on the support of $F$.

In~\cite{Fredenhagen:2013cna} the relation between the free and interacting time evolution has been analysed. The free time evolution corresponds to a translation of the support functions. Indicating by $F_f$ the local field smeared with a test function $f$, the free time evolution is such that
\[
\alpha_s(F_f) = F_{f_s}, \qquad f_s(t,\mathbf{x}):=f(t-s,\mathbf{x}).
\]
The interacting time evolution acts on $\mathcal{F}_I$ as
\[
\alpha^V_s(R_{V}(F_f)) = R_{V}(\alpha_sF_f)=R_{V}(F_{f_s}). 
\]
If $F$ is supported in $\Sigma_\epsilon$, the relation between the interacting and free time evolution is described by a cocycle $U_V(t)$ such that 
\[
\alpha^V_t(F) = U_V^* \star \alpha_t(F) \star U_V
\]
In~\cite{Fredenhagen:2013cna} it has been shown how to construct the cocycle $U_V$ as a Dyson series of products of its generator $K_V$, see equation (25) in~\cite{Fredenhagen:2013cna}
where, for any $V = \int \chi h\mathcal{L}_{I}$, 
\[
K_V = R_V(\mathcal{H}), \qquad   \mathcal{H} = \int \dot{\chi} h \mathcal{L}_{I}.
\]
Having  this cocycle at our disposal it is possible to construct the equilibrium (KMS) state repeating the  so-called Araki-construction also for interacting field theories, when $h$ is of compact support. In particular it holds that  
\[
\omega_\beta^V(A) = \frac{\omega(A\star U_V(i\beta))}{U_V(i\beta)}.
\]
The state constructed in this way depends explicitly on $h$, however, the limit $h\to1$ taken in the sense of van Hove is well defined, it is independent on $\chi$ and it constitutes an equilibrium state for the interacting theory in the adiabatic limit~\cite{Lindner:2013ila,Fredenhagen:2013cna} .

 The methods developed in this project have opened the path to extending many interesting results of quantum statistical mechanics to quantum field theory.
For example, it was possible to prove return to equilibrium and stability for interaction Lagrangians of compact spatial support 
\[
\lim_{t\to\infty} \omega^\beta\circ\alpha_t^V = \omega^{\beta,V}
\]
see~\cite{Drago:2016zlf}. In that paper it is also shown that the return to equilibrium ceases to hold if the adiabatic limit is considered before the large-time limit.

% ****************************************************************************
% BIBLIOGRAPHY AREA
% ****************************************************************************

\begin{footnotesize}

\bibliographystyle{sfb676}
\bibliography{sfb676_C7}

\end{footnotesize}

% ****************************************************************************
% END OF BIBLIOGRAPHY AREA
% ****************************************************************************

\end{document}